\documentstyle[twoside,psfig]{article}

\catcode`\@=11
\long\def\@makefntext#1{
\protect\noindent \hbox to 3.2pt {\hskip-.9pt  
$^{{\eightrm\@thefnmark}}$\hfil}#1\hfill}		

\def\thefootnote{\fnsymbol{footnote}}
\def\@makefnmark{\hbox to 0pt{$^{\@thefnmark}$\hss}}	
	
\def\ps@myheadings{\let\@mkboth\@gobbletwo
\def\@oddhead{\hbox{}
\rightmark\hfil\eightrm\thepage}   
\def\@oddfoot{}\def\@evenhead{\eightrm\thepage\hfil
\leftmark\hbox{}}\def\@evenfoot{}
\def\sectionmark##1{}\def\subsectionmark##1{}}

\oddsidemargin=\evensidemargin
\renewcommand{\thefootnote}{\fnsymbol{footnote}}

\newcounter{sectionc}\newcounter{subsectionc}\newcounter{subsubsectionc}
\renewcommand{\section}[1] {\vspace{12pt}\addtocounter{sectionc}{1} 
\setcounter{subsectionc}{0}\setcounter{subsubsectionc}{0}\noindent 
	{\tenbf\thesectionc. #1}\par\vspace{5pt}}
\renewcommand{\subsection}[1] {\vspace{12pt}\addtocounter{subsectionc}{1} 
	\setcounter{subsubsectionc}{0}\noindent 
	{\bf\thesectionc.\thesubsectionc. {\kern1pt \bfit #1}}\par\vspace{5pt}}
\renewcommand{\subsubsection}[1] {\vspace{12pt}\addtocounter{subsubsectionc}{1}
	\noindent{\tenrm\thesectionc.\thesubsectionc.\thesubsubsectionc.
	{\kern1pt \tenit #1}}\par\vspace{5pt}}
\newcommand{\nonumsection}[1] {\vspace{12pt}\noindent{\tenbf #1}
	\par\vspace{5pt}}

\newcounter{appendixc}
\newcounter{subappendixc}[appendixc]
\newcounter{subsubappendixc}[subappendixc]
\renewcommand{\thesubappendixc}{\Alph{appendixc}.\arabic{subappendixc}}
\renewcommand{\thesubsubappendixc}
	{\Alph{appendixc}.\arabic{subappendixc}.\arabic{subsubappendixc}}

\renewcommand{\appendix}[1] {\vspace{12pt}
        \refstepcounter{appendixc}
        \setcounter{figure}{0}
        \setcounter{table}{0}
        \setcounter{lemma}{0}
        \setcounter{theorem}{0}
        \setcounter{corollary}{0}
        \setcounter{definition}{0}
        \setcounter{equation}{0}
        \renewcommand{\thefigure}{\Alph{appendixc}.\arabic{figure}}
        \renewcommand{\thetable}{\Alph{appendixc}.\arabic{table}}
        \renewcommand{\theappendixc}{\Alph{appendixc}}
        \renewcommand{\thelemma}{\Alph{appendixc}.\arabic{lemma}}
        \renewcommand{\thetheorem}{\Alph{appendixc}.\arabic{theorem}}
        \renewcommand{\thedefinition}{\Alph{appendixc}.\arabic{definition}}
        \renewcommand{\thecorollary}{\Alph{appendixc}.\arabic{corollary}}
        \renewcommand{\theequation}{\Alph{appendixc}.\arabic{equation}}
        \noindent{\tenbf Appendix \theappendixc #1}\par\vspace{5pt}}
\newcommand{\subappendix}[1] {\vspace{12pt}
        \refstepcounter{subappendixc}
        \noindent{\bf Appendix \thesubappendixc. {\kern1pt \bfit #1}}
	\par\vspace{5pt}}
\newcommand{\subsubappendix}[1] {\vspace{12pt}
        \refstepcounter{subsubappendixc}
        \noindent{\rm Appendix \thesubsubappendixc. {\kern1pt \tenit #1}}
	\par\vspace{5pt}}

\newcommand{\textlineskip}{\baselineskip=13pt}
\newcommand{\smalllineskip}{\baselineskip=10pt}

\def\eightcirc{
\begin{picture}(0,0)
\put(4.4,1.8){\circle{6.5}}
\end{picture}}
\def\eightcopyright{\eightcirc\kern2.7pt\hbox{\eightrm c}} 

\newcommand{\copyrightheading}[1]
	{\vspace*{-2.5cm}\smalllineskip{\flushleft
	{\footnotesize Modern Physics Letters A, #1}\\
	{\footnotesize $\eightcopyright$\, World Scientific Publishing
	 Company}\\
	 }}

\newcommand{\publisher}[2]{{\begin{center}\footnotesize\smalllineskip 
	Received #1\\
	Revised #2
	\end{center}
	}}

\def\abstracts#1#2#3{{
	\centering{\begin{minipage}{4.5in}\footnotesize\baselineskip=10pt
	\parindent=0pt #1\par 
	\parindent=15pt #2\par
	\parindent=15pt #3
	\end{minipage}}\par}} 

\newcommand{\bibit}{\nineit}
\newcommand{\bibbf}{\ninebf}
\renewenvironment{thebibliography}[1]
	{\frenchspacing
	 \ninerm\baselineskip=11pt
	 \begin{list}{\arabic{enumi}.}
        {\usecounter{enumi}\setlength{\parsep}{0pt}     
	 \setlength{\leftmargin 12.7pt}{\rightmargin 0pt} 
         \setlength{\itemsep}{0pt} \settowidth
	{\labelwidth}{#1.}\sloppy}}{\end{list}}

\newcounter{itemlistc}
\newcounter{romanlistc}
\newcounter{alphlistc}
\newcounter{arabiclistc}

\newcommand{\fcaption}[1]{
        \refstepcounter{figure}
        \setbox\@tempboxa = \hbox{\footnotesize Fig.~\thefigure. #1}
        \ifdim \wd\@tempboxa > 5in
           {\begin{center}
        \parbox{5in}{\footnotesize\smalllineskip Fig.~\thefigure. #1}
            \end{center}}
        \else
             {\begin{center}
             {\footnotesize Fig.~\thefigure. #1}
              \end{center}}
        \fi}

\newcommand{\tcaption}[1]{
        \refstepcounter{table}
        \setbox\@tempboxa = \hbox{\footnotesize Table~\thetable. #1}
        \ifdim \wd\@tempboxa > 5in
           {\begin{center}
        \parbox{5in}{\footnotesize\smalllineskip Table~\thetable. #1}
            \end{center}}
        \else
             {\begin{center}
             {\footnotesize Table~\thetable. #1}
              \end{center}}
        \fi}

\def\@citex[#1]#2{\if@filesw\immediate\write\@auxout
	{\string\citation{#2}}\fi
\def\@citea{}\@cite{\@for\@citeb:=#2\do
	{\@citea\def\@citea{,}\@ifundefined
	{b@\@citeb}{{\bf ?}\@warning
	{Citation `\@citeb' on page \thepage \space undefined}}
	{\csname b@\@citeb\endcsname}}}{#1}}

\newif\if@cghi
\def\cite{\@cghitrue\@ifnextchar [{\@tempswatrue
	\@citex}{\@tempswafalse\@citex[]}}
\def\citelow{\@cghifalse\@ifnextchar [{\@tempswatrue
	\@citex}{\@tempswafalse\@citex[]}}
\def\@cite#1#2{{$\null^{#1}$\if@tempswa\typeout
	{IJCGA warning: optional citation argument 
	ignored: `#2'} \fi}}

\def\pmb#1{\setbox0=\hbox{#1}
	\kern-.025em\copy0\kern-\wd0
	\kern.05em\copy0\kern-\wd0
	\kern-.025em\raise.0433em\box0}

\def\fnm#1{$^{\mbox{\scriptsize #1}}$}
\def\fnt#1#2{\footnotetext{\kern-.3em
	{$^{\mbox{\scriptsize #1}}$}{#2}}}

\def\runninghead#1#2{\pagestyle{myheadings}
\markboth{{\protect\footnotesize\it{\quad #1}}\hfill}
{\hfill{\protect\footnotesize\it{#2\quad}}}}
\font\tenrm=cmr10
\font\tenit=cmti10 
\font\tenbf=cmbx10
\font\bfit=cmbxti10 at 10pt
\font\ninerm=cmr9
\font\nineit=cmti9
\font\ninebf=cmbx9
\font\eightrm=cmr8

\def\qed{\hbox{${\vcenter{\vbox{			
   \hrule height 0.4pt\hbox{\vrule width 0.4pt height 6pt
   \kern5pt\vrule width 0.4pt}\hrule height 0.4pt}}}$}}

\renewcommand{\thefootnote}{\fnsymbol{footnote}} 

\def\QG{{\rm QG}}
\def\N{{\rm N}}
\def\P{{\rm P}}
\begin{document}
\setlength{\textheight}{7.7truein}  

\thispagestyle{empty}
\markboth{\protect{\footnotesize\it Possible Astrophysical Probes of
Quantum Gravity}}{\protect{\footnotesize\it Possible Astrophysical
Probes of Quantum Gravity}} \runninghead{Possible Astrophysical
Probes of Quantum Gravity}{Possible Astrophysical Probes of Quantum
Gravity}

\normalsize\textlineskip

\setcounter{page}{1}
\copyrightheading{}			
\vspace*{0.88truein}

\centerline{\bf POSSIBLE ASTROPHYSICAL PROBES OF QUANTUM GRAVITY}
\vspace*{0.4truein}
\centerline{\footnotesize SUBIR SARKAR}
\baselineskip=12pt
\centerline{\footnotesize\it Theoretical Physics, University of
Oxford, 1 Keble Road}
\baselineskip=10pt
\centerline{\footnotesize\it Oxford OX1 3NP, United Kingdom}
\vspace*{0.228truein}

\publisher{(received date)}{(revised date)}

\vspace*{0.23truein}
\abstracts{A satisfactory theory of quantum gravity will very likely
require modification of our classical perception of space-time,
perhaps by giving it a `foamy' structure at scales of order the Planck
length. This is expected to modify the propagation of photons and
other relativistic particles such as neutrinos, such that they will
experience a non-trivial refractive index even {\em in vacuo}. The
implied spontaneous violation of Lorentz invariance may also result in
alterations of kinematical thresholds for key astrophysical processes
involving high energy cosmic radiation. We discuss experimental probes
of these possible manifestations of the fundamental quantum nature of
space-time using observations of distant astrophysical sources such as
gamma-ray bursts and active galactic nuclei.}{}{}


\vspace*{2pt}

\medskip
\noindent
{\small {\it ``We finally turn to the most daunting problem of quantum
gravity, the nearly complete lack of observational and experimental
evidence that could point us in the right direction or provide tests
for our models. The ultimate measure of any theory is its agreement
with Nature; if we do not have any such tests, how will we know
whether we are right?''} 

\smallskip 
\hfill S. Carlip (2001)\cite{carlip}}


\baselineskip=13pt	        
\normalsize              	
\section{Introduction}		
\vspace*{-0.5pt}
\noindent
Quantum Mechanics, Special Relativity and General Relativity
correspond to three vertices of Bronshtein's `cube of theories'
constructed on axes labelled by the three dimensionful constants $c$,
$\hbar$ and $G_\N$, which define the fundamental units of length, mass
and time.\cite{cube} Each of these theories has been extremely
successful in describing Nature, yet their union into a consistent
(and comprehensible!) theory of quantum gravity still eludes
us. Moreover it is often stated that the quantum effects of gravity
may never be accessible to experiment because they would be manifest
only on the tiny Planck length scale,
$l_\P\equiv\sqrt{\hbar\,G_\N/c^3}\simeq1.6\times10^{-33}$~cm
(corresponding to the huge energy scale
$M_\P\equiv\sqrt{\hbar\,c/G_\N}\simeq2.2\times10^{-8}$~g or
$1.2\times10^{19}$~GeV in units with $c=\hbar=1$). However, gravity,
being a non-renormalisable interaction in the language of quantum
field theory, can perhaps leave a distinctive imprint at energies much
lower than the Planck scale if it violates some fundamental symmetry
of the effective low energy theory, resulting in qualitatively new
phenomena (akin to the violation of parity in nuclear radioactive
decay, at energies far below the true scale of the responsible weak
interaction).

At present there are two main approaches to the quantisation of
continuum gravity.\cite{carlip} The first is the canonical programme
for a non-perturbative and background-independent quantisation of
general relativity, viz. the loop-gravity formalism.\cite{loopqg} The
second is superstring theory,\cite{superstrings} and its
non-perturbative D(irichlet)-brane extension.\cite{strings} Apart from
these formal mathematical approaches there have been other discussions
of the likely physical manifestations of quantum gravity, e.g of the
possibility that quantum space-time has a `foamy'
structure,\cite{stfoam} in which Planck-size topological fluctuations
resembling black holes with microscopic event horizons appear
spontaneously out of the vacuum and subsequently evaporate back into
it. These black-hole horizons have been viewed as providing an
`environment' that might induce quantum decoherence of apparently
isolated matter systems.\cite{ehns,zurek} In this picture Lorentz
invariance (LI) would appear to be lost in the splitting between the
matter system and the quantum-gravitational `environment'; such a
breaking of LI can be considered a property of the
quantum-gravitational ground state, and therefore a variety of {\em
spontaneous} breaking. This also occurs in the loop-gravity formalism,
as demonstrated in the semi-classical limit.\cite{gp,amu} In critical
string theory there is no Planck scale space-time foam but in {\em
non-critical} `Liouville' string theory,\cite{liouville} the link can
be made by viewing D-branes as space-time defects, giving rise to a
cellular structure in the space-time manifold;\cite{emnfoam,emnfluc}
Thus it seems not unreasonable to expect spontaneous violation of LI
at high energies as a {\em generic} signature of quantum gravitational
effects. Although a comprehensive theoretical framework may be lacking
at present, such an exciting possibility is adequate motivation for a
phenomenological approach to the question of whether such tiny
violations can be observed.\cite{aemns,rev}

\setcounter{footnote}{0}
\renewcommand{\thefootnote}{\alph{footnote}}

\vspace*{2pt}
\section{Quantum gravity and possible modifications to dispersion relations}
\noindent
In this talk I will discuss whether astrophysical observations can
test possible LI violations due to the quantum structure of
space-time. We are interested in signatures that are characterised by
deviations from conventional quantum field theory, which would
presumably be suppressed by some power of the Planck
mass.\fnm{a}\fnt{a}{Our approach is somewhat different from identifying all
possible LI violating terms which can be added to the Standard Model
at the {\em renormalisable} level; the possible phenomenology of such
terms and current experimental bounds on them have been discussed in
detail elsewhere.\cite{ck,cg}} It appears that several such effects
are at the edge of observability if the suppression is by just a {\em
single} power of $M_\P$.

This may well be the case for the possible effects of a
quantum-gravitational environment on the propagation of a massless
particle such as a photon. At energies small compared to the Planck
scale a series expansion of the usual dispersion relation in the small
parameter $E/M_\P$ is justified, i.e. \cite{aemns}
\begin{equation}
 c^2 p^2 = E^2 \left[1 + \xi\left(\frac{E}{M_\P}\right) 
           + {\cal O}\left(\frac{E^2}{M_\P^2}\right) + \ldots \right].
\label{disp}
\end{equation}
Such a relation with $\xi=+1$ is consistent with studies
\cite{kappapoincare} of quantum deformations of space-time symmetries
(viz. the Hopf extension of Poincar\'e algebra), if we identify the
`deformation parameter' $\kappa$ as the inverse of the Planck length
and impose rotational symmetry (i.e. invariance under the O(3)
subgroup of the SO(3,1) Lorentz group).\cite{kppheno}\fnm{b}\fnt{b}{A
possible astrophysical test of $\kappa$-Poincar\'e deformations had
been suggested earlier.\cite{dk}} Such a deformed dispersion relation
also arises in non-critical Liouville string theory, as has been
demonstrated using the picture of D-brane recoil.\cite{emnfoam} This
is essentially because the action for such theories is proportional to
the {\em square root} of the string tension (which has dimensions of
the inverse of the fundamental length {\em squared}); by contrast in
critical string theory the action is proportional to the string
tension, consequently only corrections of ${\cal O}(E^2/M_\P^2)$ would
be expected.\cite{sumit}\fnm{c}\fnt{c}{Such $E^2/M_\P^2$ corrections are
also found for the dispersion relation of transversely polarised
photons due to graviton induced vacuum polarisation in 3+1
dimensions.\cite{bcfmg}} Given however that the connection between
critical string theory (in 10 dimensions) and our low energy world in
3+1 dimensions is still an open question, it would seem premature to
dismiss the possibility of a dispersion term of ${\cal O}(E/M_\P)$. It
is more interesting to ask if we can test for such a term {\em
experimentally} to obtain a constraint on theories. Somewhat
unexpectedly this turns out to be possible.\cite{aemns}

\subsection{Time-of-flight studies of radiation from distant objects}
\noindent
Assuming the usual Hamiltonian equation of motion
$\dot{x_i}=\partial\,H/\partial\,p_i$ to still be valid, the above
dispersion relation (\ref{disp}) implies an {\em energy dependent}
speed of light,
\begin{equation}
 v = \frac{\partial E}{\partial p} \simeq
 c\left[1 - \xi\left(\frac{E}{M_\P}\right)\right].
\label{refractive}\end{equation}
We observe that the effect (\ref{refractive}) is easy to distinguish
from dispersion in any field theoretical vacuum or plasma, which
always {\em decreases} with increasing energy.\cite{ftdisp} Thus when
we study a source at distance $L$, we will see a quantum gravity
induced {\em time delay} (for $\xi$ +ve) in the arrival of the more
energetic photons as compared to the less energetic ones (differing in
energy by an amount $\Delta E$) of order:
\begin{equation}
 \Delta t \simeq \xi \frac{L}{c} \times \frac{\Delta E}{M_\P}.
\label{delay} 
\end{equation}
An analogous effect arises in the semi-classical limit of loop
gravity, if the gravitational degrees of freedom are assumed to be in
a `weave' state.\cite{gp}

The energy-dependent refractive index (\ref{refractive}) would induce
dispersion in the arrival times of photons emitted in a short pulse by
an astrophysical source. It is also possible that different photons
with the same energy (frequency) might travel at different velocities,
as is suggested by higher-order studies in the D-brane approach to
quantum gravity.\cite{emnfluc} This would provide a second possible
source of dispersion in a wave packet, beyond that associated with
differing frequencies:
\begin{equation}
 \delta t \sim \frac{\sqrt{\langle\sigma^2\rangle}}{L},  
\label{fluc} 
\end{equation}
where $\sigma^2$ denotes a quantum fluctuation about the specific
classical background. Both these effects can be described in terms of
quantum fluctuations in the light cone.\cite{lcfluc,emnfluc} Moreover
if the weave states of loop quantum gravity have definite parity then
one also expects different propagation velocities for left- and
right-handed polarised photons, viz. bi-refringence of the
vacuum.\cite{gp} This possibility is however severely constrained
already by observations of distant radio galaxies which exhibit
detectable linear polarisation in their emissions;\cite{gk} the
corresponding time delays for high energy photons in this model may
then be too small to be observable. However a study of the propagation
of spin 1/2 fermions in the semi-classical limit of loop quantum
gravity shows that there would still be detectable time delays between
neutrinos and photons from GRBs.\cite{amu}\fnm{d}\fnt{d}{Such a time delay
is also found \cite{bc} for the LI violating extension of the Standard
Model;\cite{ck} this may not however be observable if the pulses are
smeared out due to the dispersive effect (\ref{fluc}).\cite{emngzk}}

\begin{figure}[htbp] 
\vspace*{13pt}
\centerline{\psfig{angle=-90,width=5in,file=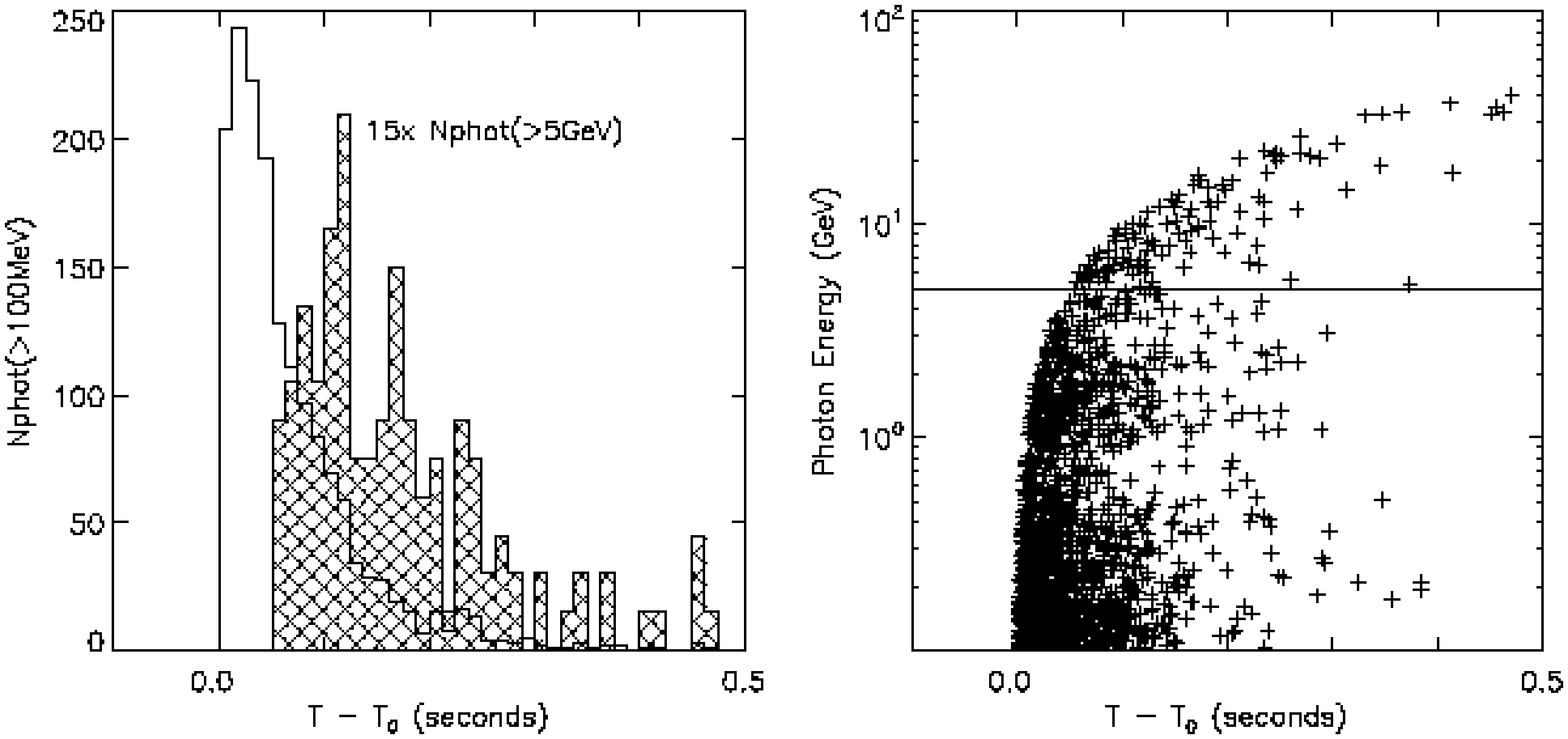}} 
\vspace*{13pt}
\fcaption{The expected sensitivity of GLAST to the quantum
 gravity induced dispersion in Eq.(\protect\ref{delay}) using
 observations of gamma-ray bursts.\protect\cite{glast} The left
 panel shows the simulated composite pulse from 20 bright GRBs observed
 over 2 years, with energy threshold of 0.1 GeV (open histogram) and 5
 GeV (cross-hatched histogram). The right panel shows the arrival time
 as a function of the photon energy; the expected dispersion is
 clearly seen above a 5 GeV threshold (horizontal line).}
\label{glastdisp}
\end{figure}

The figure of merit for such tests is
\begin{equation} 
 M_\QG \equiv \xi \frac{LE}{c \Delta t},
\label{figmerit}
\end{equation}
where $M_\QG$ is the highest energy scale that can be probed using a
source of photons of energy $E$ at distance $L$ which exhibits
distinct structure on a time scale of order $\Delta\,t$. As we
emphasised,\cite{aemns} gamma-ray bursts (GRBs) have particularly
large figures of merit. Some of them exhibit microstructures in their
light curves of a millisecond or less, they are likely to emit
$\gamma$ rays in the GeV or even TeV range, and many are now
definitely known to be located at cosmological distances.\cite{grbrev}
We estimated that GRB observations might already be sensitive to a
quantum-gravity scale $M_{\QG}\sim10^{16}$~GeV, and suggested that
atmospheric \v{C}erenkov telescopes (ACTs) would be able to improve on
this sensitivity. The Whipple ACT group has applied this idea to
observations of the active galactic nucleus (AGN) Mkn~421,
establishing a lower limit $M_{\QG}>4\times10^{16}$~GeV.\cite{whipple}
The potential of next-generation ACTs (such as VERITAS, MAGIC and
HESS) for improving such limits has been assessed.\cite{act} A similar
lower limit has been inferred from observations of gamma-ray
pulsars.\cite{kaaret} A careful analysis of present data on GRBs
however finds a slightly weaker limit
$M_{\QG}>10^{15}$~GeV,\cite{efmmn}\fnm{e}\fnt{e}{A more severe limit of
$M_{\QG}>8.3\times10^{16}$~GeV was claimed to follow from observations
of GRB 930131;\cite{schaefer} however this object has no measured
redshift, hence a very uncertain distance.} but remarkably enough,
space-borne experiments such as AMS and, particularly, GLAST are
expected to definitively find (or rule out) the expected time delay
for $M_{\QG}\sim\,M_\P$. Figure~\ref{glastdisp} demonstrates that
GLAST (scheduled for launch in 2005) will be able to do this in just 2
years of operation!

\setcounter{footnote}{0}
\renewcommand{\thefootnote}{\alph{footnote}}

\subsection{Modifications of relativistic kinematics due to LI violation}
\noindent
Among the key considerations for the above experiments is the question
of just how far can one see into the universe at high energies; for
example at TeV energies, $\gamma$-rays are expected to be severely
attenuated through pair-production on the intervening cosmic infra-red
background (CIB) radiation (above a threshold energy $m_e^2/E_{\rm
CIB}\sim10$~TeV) so that ACTs should see an exponential cutoff in AGN
spectra at such energies.\cite{atten} The absence of the expected
cutoff in the spectra of the relatively nearby sources Mkn~421
($z=0.031$) and Mkn~501 ($z=0.033$) was thus interpreted as setting
limits on the (otherwise poorly known) intensity of the
CIB.\cite{atten2} This radiation is created by stars and reprocessed
by dust so its intensity is sensitive to the entire history of galaxy
formation and can be estimated in specific models.\cite{pbsm}

\begin{figure}[htbp] 
\vspace*{13pt}
\centerline{\psfig{height=2.25in,file=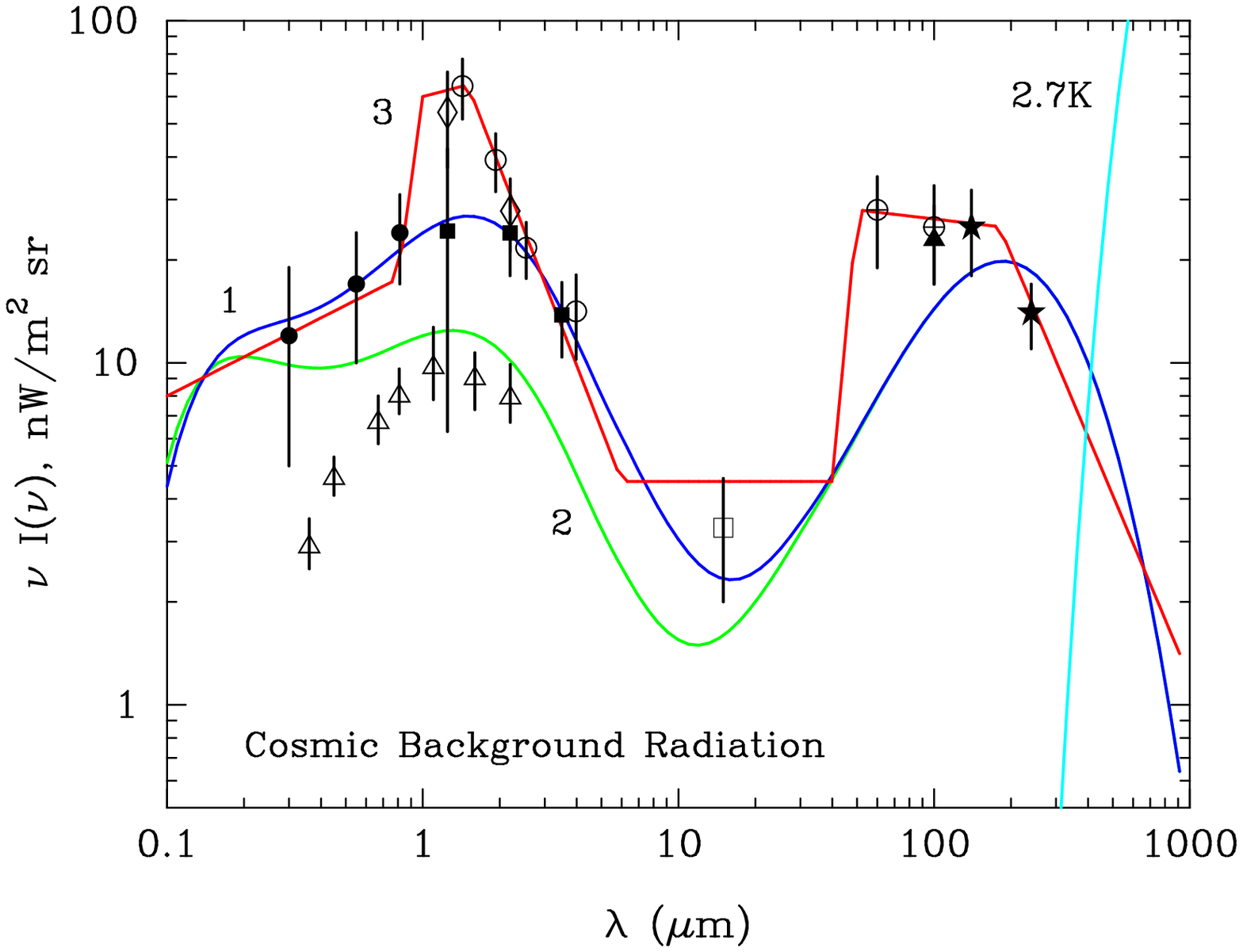}
            \psfig{width=2.0in,file=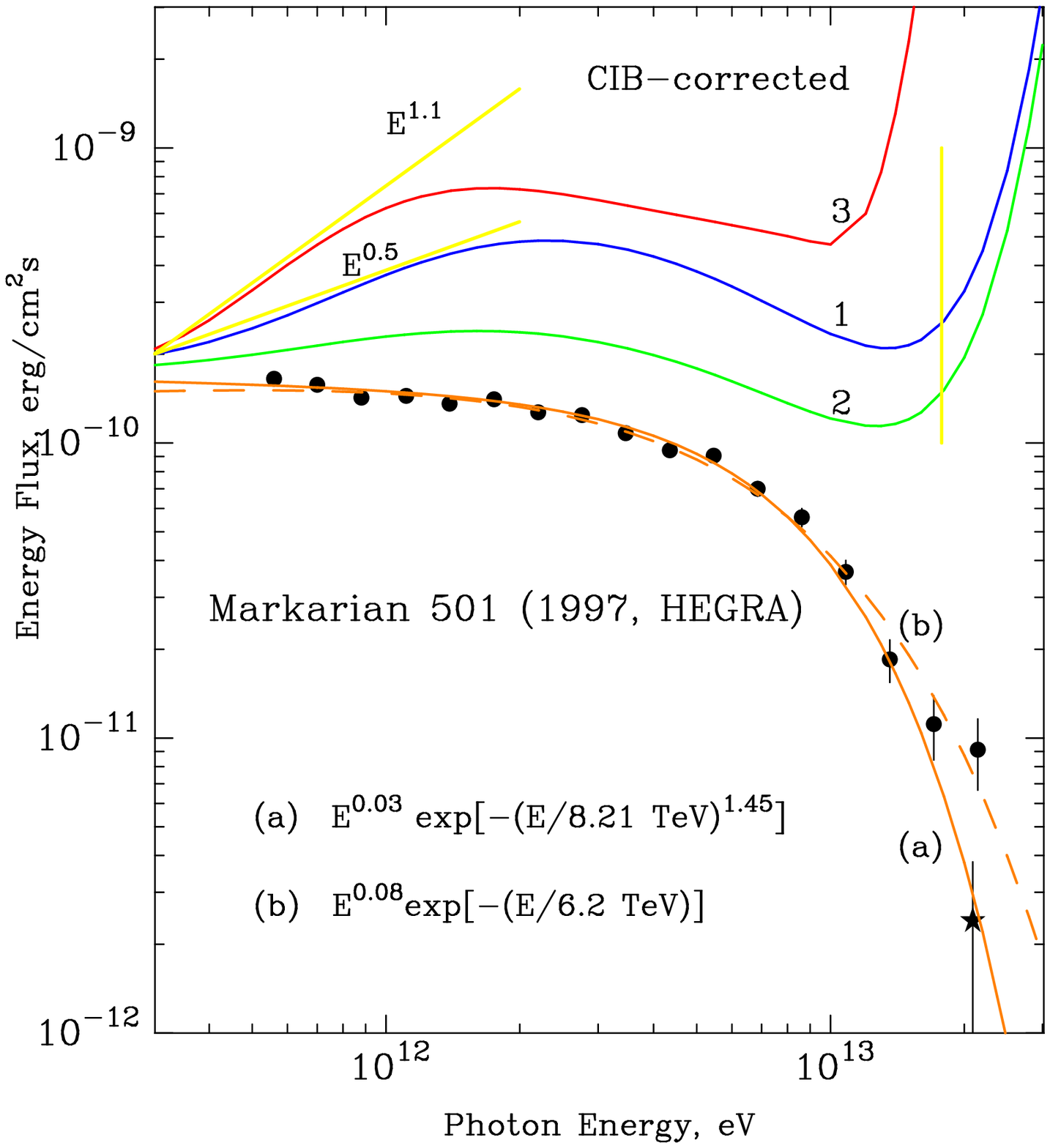}} 
\vspace*{13pt} 
\fcaption{Reconstruction \protect\cite{felix} of the emission spectrum
 of Mkn~501 from TeV $\gamma$-ray observations (right
 panel),\protect\cite{hegra,hegra2} after allowing for attenuation by
 the observed cosmic infrared background (left panel). The vertical
 line at 17 TeV indicates the point upto which the spectrum is mesured
 reliably.}
\label{irgamma}
\end{figure}

When the first direct measurements of the CIB were made, notably by
the DIRBE instrument on COBE,\cite{cib} the fluxes proved surprisingly
to be {\em above} the limits set by the above argument. It had been
pointed out however \cite{kluzniak,kifune,abgg} that the LI violation
implicit in the dispersion relationship (\ref{disp}) would affect the
kinematics of the attenuation process $\gamma\gamma\to\,e^+e^-$ such
that there is {\em no} physical threshold for this process, hence no
attenuation of TeV energy $\gamma$-rays even from very distant
sources, independently of the intensity of the
CIB.\fnm{f}\fnt{f}{There is in fact no change in the threshold energy
condition for the $\kappa$-deformed Poincar\'e
algebra,\cite{kappapoincare} but there are significant corrections to
the kinematics above threshold.\cite{ap,kpthresh}} This possibility
was promoted as a solution of the `IR-TeV $\gamma$-ray
crisis'\cite{pm} created by the non-observation of a sharp cutoff in
the spectrum of Mkn~501 upto $\sim21$~TeV by the HEGRA
ACT.\cite{hegra} Their point is made in Figure~\ref{irgamma} which
shows an updated version of this argument.\cite{felix} The left panel
shows a compendium of CIB observations with 3 reference curves drawn
to indicate the range of possibilities --- curve 1 is a `nominal' fit,
curve 2 is close to the expectations from modelling of galaxy
formation,\cite{pbsm} while the `extreme' curve 3 is drawn to match
the highest CIB fluxes reported (which may have suffered contamination
from zodiacal light). The right panel shows the intrinsic $\gamma$-ray
spectrum of Mkn~501 reconstructed from the HEGRA
observations.\cite{hegra} Although the observed spectrum does bend
down, particularly after reducing the intensity in the highest energy
bin (down to the starred point) at 21.45 TeV to allow for possible
spillover from lower energies,\cite{hegra2} making allowance for the
expected attenuation by the CIB (especially at $\sim100\,\mu$m) still
requires a rather unphysical turnup in the intrinsic spectrum at
$\sim10$~TeV.\cite{pm} This is so even for the `nominal' fit (curve
1), although the turnup is not as sharp as when the `extreme' fit
(curve 3) is used. There is no anomaly if curve 2 reflects the true
CIB and this is indeed close to estimates based on galaxy evolution
models.\cite{ms,joel} However such estimates are necessarily
model-dependent and have many uncertainties so it seems unreasonable
to conclude on this basis alone that there is no
problem.\cite{sg,venya} Clearly the issue will be resolved with
further observations of even more distant AGN, e.g. H1426+428
($z=0.129$); present observations are in fact consistent with
attenuation by the CIB with {\em no} need for LI
violation.\cite{hegra3}

Similarly it has been noted that photopion production of ultra high
energy cosmic rays (UHECR) on the cosmic microwave background (CMB),
$p\gamma\to\Delta(1232)\to\pi^+\ldots$, which ought to result in a
`GZK' cutoff \cite{gzk} of their energy spectrum above
$\sim5\times10^{10}$~GeV (if they are indeed protons of extragalactic
origin), may be evaded by LI violation at high
energies.\cite{kc,st,gm,cg2,kifune} This is an interesting possibility
given that the UHECR spectrum is seen to extend well beyond the GZK
cutoff while their sky distribution is consistent with isotropy,
suggestive of cosmologically distant sources.\cite{uhecr} However
there are plausible alternative explanations for this, for example
such UHECRs may originate {\em locally} in the halo of our Galaxy,
thus evading GZK energy losses while still being
quasi-isotropic.\cite{venya,me} The ongoing Pierre Auger experiment
will soon provide improved determinations of the composition, spectral
shape and anisotropy of UHECRs which will critically test such
explanations.

The existence of `threshold anomalies' is thus rather controversial
from the observational point of view. As a cautionary theoretical
note, the assumption of energy conservation implicit in recent
analyses of scattering kinematics \cite{kluzniak,kifune,abgg,ap,ap2}
may not in fact be valid --- e.g. in the space-time foam picture
energy is conserved only in a {\em statistical}
sense.\cite{ehns,emngzk,nv} In this picture there is no connection
between the TeV $\gamma$-ray and the UHECR puzzles which require
e.g. very different values of the LI violating parameter.\cite{nlov}
This is in contrast to the claim \cite{ap2,gac} that both of these
puzzles, as well as an alleged `anomaly' in the development of cosmic
ray air showers,\cite{shower}\fnm{g}\fnt{g}{It is stated \cite{gac}
that the observed development of cosmic ray air showers can be
explained ``by assuming that ultra-high-energy neutral pions are much
more stable than low energy ones''. However the cited source
\cite{shower} does not claim any such anomaly; the authors simply
obtain a bound on possible LI violating effects.\cite{cg} A subsequent
analysis \cite{vs} notes on the contrary that the dispersion effect
(\ref{disp}) would imply {\em unacceptable} changes in the development
of the em cascade --- however this conclusion is based on assuming
energy conservation which may be invalid as noted above.} can all be
traced to the same quantum gravity induced dispersion effect
(\ref{disp}). This underlines the limitations of a purely
`phenomenological' approach --- without explicit guidance from a
theory of quantum gravity, it is difficult to be certain of the
effects of possible LI violation in different experimental contexts.

\begin{figure}[htbp] 
\vspace*{13pt}
\centerline{\psfig{width=2.7in,file=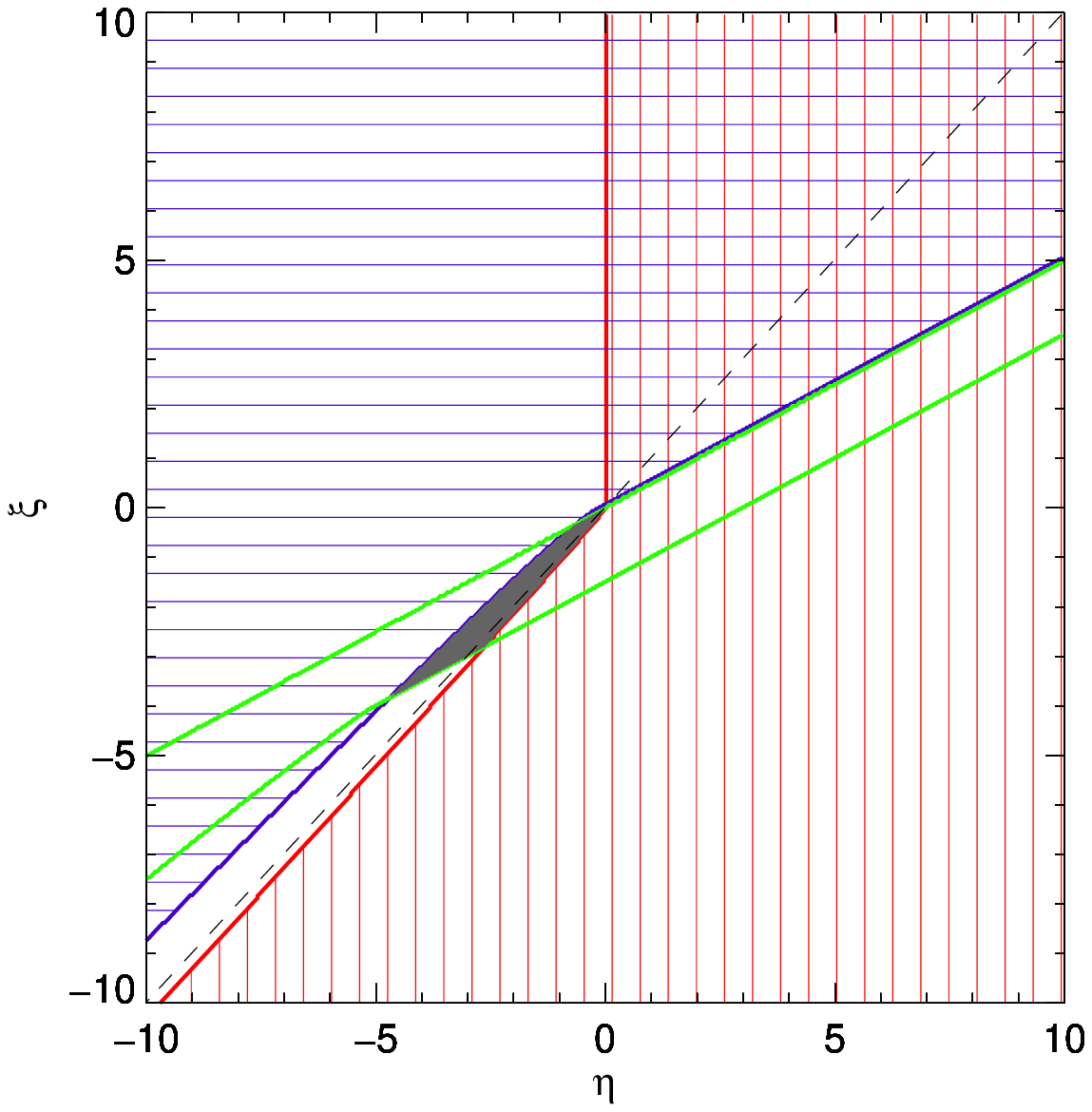}} 
\vspace*{13pt} 
\fcaption{Constraints \protect\cite{ljm} on the LI violating parameters
 for photons ($\xi$) and electrons ($\eta$) (defined as in
 Eq.(\protect\ref{disp}) but with {\em opposite} sign). The
 horizontally ruled region is excluded by the absence of vacuum photon
 decay and the vertically ruled region by the absence of vacuum
 \v{C}erenkov radiation. The threshold for attenuation of cosmic
 $\gamma$-rays on CIB photons (of energy 0.025 eV) is raised from 
 10 to 20 TeV in moving from the upper diagonal line to the lower
 one. The dashed line is $\xi=\eta$.}
\label{constraints}
\end{figure}

Alternatively one can simply parameterise every possible LI violating
term that can be added to the low energy effective theory, with LI
assumed to hold in the `preferred' frame identified by the (absence of
a) dipole anisotropy in the CMB.\cite{cg} In such schemes, particles
in any other frame may move faster or slower than light (in contrast
to the space-time foam picture where only {\em subluminal} motion is
possible \cite{emnfluc}). This suggests other interesting
astrophysical tests e.g. if the maximum attainable speed of photons
$c_\gamma$ exceeds the maximum attainable speed of electrons $c_e$
then photons of high enough energy would be unstable against decay
$\gamma\to\,e^+e^-$; if the reverse were true then one can have
\v{C}erenkov radiation $e\to\,e\gamma$ {\em in vacuo}.\cite{cg3,cg}
Since we do observe photons of energy upto $\sim20$~TeV from distant
AGN \cite{hegra} and infer the existence of electrons with energies
upto $\sim100$~TeV (from observations of non-thermal X-rays and TeV
$\gamma$-rays in some supernova remnants,\cite{snr} restrictive limits
can be set on such LI violating effects, in addition to those
discussed earlier concerning possible `threshold anomalies'
\cite{cg2,cg3,cg} and pion stability.\cite{shower} A systematic
analysis of such constraints has recently been undertaken
\cite{ljm,km} in the framework of the quantum gravity inspired
dispersion relation (\ref{disp}), but allowing for a different value
of $\xi$ for photons than for other particles (the correspondance to
the alternative LI violating formalism \cite{cg} can be made
\cite{kifune} by setting $\xi_i\,E/M_\P$ equal to
$c_\gamma^2-c_i^2$). These analyses deduce joint constraints on the LI
violating parameters from a careful examination of the relativistic
kinematics of all `forbidden' processes (assuming energy conservation)
and find that a large region of parameter space is ruled out as seen
in Figure~\ref{constraints}.\cite{ljm} This shows the allowed region
in $\xi-\eta$ space, where $\xi$ is the same quantity as defined for
photons in Eq.(\ref{disp}) but now defined with {\em opposite} sign,
and $\eta$ is defined similarly for electrons. Thus Eq.(\ref{disp})
would suggest $\xi=\eta=-1$ as has been considered in previous
work.\cite{aemns,kluzniak,kifune,pm} We see that this is {\em not}
ruled out, which is unsurprising since in this case there is no
`superluminal' motion.

\section{Conclusions}
\noindent
Our modest proposal \cite{aemns} that a possible effect of quantum
gravity on the propagation of photons may be amenable to observation,
has triggered quite a burst of activity and been generally well
received by theorists.\cite{qg} However, as already remarked, one
needs a concrete formalism to do further work, in particular to study
the possible effects of the associated spontaneous LI violation for
relativistic kinematics. So far this has been done mainly in the
framework of the Liouville string theory model of space-time
foam,\cite{emnfluc,emngzk} the semi-classical limit of loop quantum
gravity,\cite{gp,amu,ap'} the deformed $\kappa$-Poincar\'e
algebra,\cite{kppheno,kpthresh}\fnm{h}\fnt{h}{The recent `Doubly
Special Relativity' constructions \cite{dsr1,dsr2} also fall in this
category.\cite{dsrkp}} and in a model for quantum energy-momentum
uncertainties in space-time foam.\cite{nlov} As these unconventional
ideas attract increasing attention in the community, new
phenomenological tests are being suggested. For example it has been
noted that if energetic particles can travel faster than gravitons
then they would emit {\em gravitational} \v{C}erenkov radiation at an
unacceptable rate.\cite{mn} Moreover the dispersive effects of
space-time foam, if {\em flavour dependent}, would induce neutrino
oscillations with parameters in conflict with experiment.\cite{bef}
The associated light cone fluctuations may also distort the observed
Planckian spectrum of the CMB.\cite{lcfluc2} The non-critical
Liouville string model has been shown to escape these
constraints,\cite{emnesc} but the consequences for other models remain
to be worked out and confronted with observations. What is clear is
that there is indeed a phenomenology of quantum gravity and that this
is a most exciting time for its practitioners.

\nonumsection{Acknowledgments}
\noindent
I thank Dharamvir Ahluwalia and Naresh Dadhich for the invitation to
participate in this meeting, all the participants for making it such a
stimulating experience, and IUCAA for its generous hospitality. This
talk draws on work done with (and by) Nick Mavromatos and
collaborators, who I thank for many discussions. I acknowledge a
travel grant from the UK Particle Physics \& Astronomy Research
Council.

\nonumsection{References}

\end{document}